\documentclass[lettersize,journal]{IEEEtran}
\usepackage{amsmath,amsfonts}
\usepackage{algorithmic}
\usepackage{algorithm}
\usepackage{array}
\usepackage[caption=false,font=normalsize,labelfont=sf,textfont=sf]{subfig}
\usepackage{textcomp}
\usepackage{stfloats}
\usepackage{url}
\usepackage{verbatim}
\usepackage{graphicx}
\usepackage{cite}
\usepackage{balance}
\begin{document}

\title{SDC-Net: A Domain Adaptation Framework with Semantic-Dynamic Consistency for Cross-Subject EEG Emotion Recognition}
\author{Jiahao Tang\textsuperscript{\textdagger}, \textit{Student Member, IEEE}, Youjun Li\textsuperscript{\textdagger}, Xiangting Fan, Yangxuan Zheng, Siyuan Lu, Xueping Li, Peng Fang, Chenxi Li$^{*}$ and Zi-Gang Huang$^{*}$ 
	\thanks{\emph{*Corresponding author: ChenXi Li and Zi-Gang Huang}\\
		\indent \textsuperscript{\textdagger} These authors contributed equally to this work.}
	\thanks{This work was supported by the Natural Science Foundation of China (No. 11975178), Natural Science Basic Research Program of Shaanxi (No. 2023-JC-YB-07).}
	\thanks{Jiahao Tang, Youjun Li, Xiangting Fan, Yangxuan Zheng, Siyuan Lu and Zi-Gang Huang are affiliated with the Institute of Health and Rehabilitation Science, School of Life Science and Technology, Xi'an Jiaotong University, as well as the Research Center for Brain-Inspired Intelligence, Xi'an Jiaotong University. Chenxi Li and Peng Fang are affiliated with  Department of Military Medical Psychology, Fourth Military Medical University, Xi’an, 710032, PR China   Military Medical Psychology School, Fourth Military Medical University, Xi'an, People’s Republic of China. Xueping Li is affiliated with School of Automation and Information Engineering, Xi’an University of Technology, Xi’an, China}}
\maketitle

\textbf{Preprint Notice: This work has been submitted to the IEEE for possible publication. Copyright may be transferred without notice, after which this version may no longer be accessible.}

\begin{abstract}
Emotion recognition based on electroencephalography (EEG) holds significant promise for affective brain-computer interfaces (aBCIs). However, its practical deployment faces challenges due to the variability within inter-subject and the scarcity of labeled data in target domains. To overcome these limitations, we propose SDC-Net, a novel Semantic-Dynamic Consistency domain adaptation network for fully label-free cross-subject EEG emotion recognition. First, we introduce a Same-Subject Same-Trial Mixup strategy that generates augmented samples through intra-trial interpolation, enhancing data diversity while explicitly preserving individual identity to mitigate label ambiguity. Second, we construct a dynamic distribution alignment module within the Reproducing Kernel Hilbert Space (RKHS),  jointly aligning marginal and conditional distributions through multi-objective kernel mean embedding, and leveraging a confidence-aware pseudo-labeling strategy to ensure stable adaptation. Third, we propose a dual-domain similarity consistency learning mechanism that enforces cross-domain structural constraints based on latent pairwise similarities, facilitating semantic boundary learning without reliance on temporal synchronization or label priors. To validate the effectiveness and robustness of the proposed SDC-Net, extensive experiments are conducted on three widely used EEG benchmark datasets: SEED, SEED-IV, and FACED. Comparative results against existing unsupervised domain adaptation methods demonstrate that SDC-Net achieves state-of-the-art performance in emotion recognition under both cross-subject and cross-session conditions. This advancement significantly improves the accuracy and generalization capability of emotion decoding, laying a solid foundation for real-world applications of personalized aBCIs. The source code is available at: https://github.com/XuanSuTrum/SDC-Net.
\end{abstract}


\section{Introduction}
\label{sec:introduction}
\IEEEPARstart{E}{motion} is fundamental to human experience, reflecting the interplay between physiological states and neural activity. With advances in affective computing, emotional states are increasingly quantified as measurable variables, driving emotion-aware systems in medical, industrial, and consumer domains\cite{hsueh2023cardiogenic,anderson2014framework}. Intelligent human–machine interaction further underscores the need for nuanced emotional perception in applications such as rehabilitation robotics, health monitoring, and affective companionship\cite{khare2024emotion,jiang2025motion}. Future systems must move beyond external behaviors toward internal states to enable natural and personalized interaction.  

Emotion recognition draws on behavioral and physiological cues, including facial expressions, gestures, speech, and biosignals\cite{anderson2014framework}. Physiological signals—such as ECG, EMG, EOG, respiration, and EEG—are particularly robust due to their objectivity\cite{si2023cross}. Among them, EEG is favored for its non-invasive nature and high temporal resolution, enabling real-time emotion monitoring in brain–computer interfaces and adaptive systems. Despite progress, two challenges persist: (1) large inter-subject variability and intra-subject non-stationarity hinder generalization; (2) acquiring reliable emotion labels is labor-intensive, limiting dataset scalability. These issues restrict real-world deployment, highlighting the need for label-efficient and subject-robust methods. Transfer learning addresses these challenges by leveraging labeled data from a source domain to improve performance in an unlabeled target domain\cite{wan2021review,pan2009survey}. Domain adaptation (DA) in particular reduces inter-subject variability by aligning distributions and transferring representations between domains\cite{pan2010q}.

Existing state-of-the-art DA methods for EEG-based emotion recognition can be broadly categorized into three paradigms. The first line of work enforces consistency constraints at the output or feature level to improve cross-domain generalization. Representative approaches include multi-source feature extraction \cite{chen2021ms,li2023ms}, adversarial feature alignment (e.g., DANN \cite{li2019domain}), and pseudo-label propagation strategies \cite{zhong2025unsupervised}. While effective in aligning distributions, these approaches often overlook inter-instance semantic structures and are vulnerable to noisy pseudo-labels, especially in unsupervised target settings. To address discriminative limitations, a second category of methods introduces structural modeling techniques, such as pairwise or triplet loss constraints \cite{li2022dynamic,zhang2024evolutionary,zhou2023pr,zhou2024eegmatch} and prototype-based representations. These methods typically guide target samples toward pre-learned emotion category prototypes in the source domain, treating them as "semantic anchors." However, distributional shifts can misalign these anchors in the target domain, yielding suboptimal guidance and diminished transfer performance. More recently, contrastive learning has gained popularity for learning subject-invariant and emotion-discriminative representations by maximizing agreement between samples from the same emotional state and minimizing it across different states \cite{shen2022contrastive,dai2025contrastive,hu2025cross,wang2024gc}. Nevertheless, most contrastive frameworks rely on temporally synchronized experimental protocols to define positive and negative pairs, which implicitly leverage label information and limit their applicability in real-world unsupervised scenarios. A more comprehensive review of related studies, including deep learning, data augmentation, and domain adaptation methods for EEG-based emotion recognition, is provided in Supplementary Information (Section S5).

Despite recent advances, current methods in cross-subject EEG emotion recognition still face three fundamental challenges. First, although data augmentation strategies-such as GAN-based generation or cross-subject sample mixing\cite{zhang2022ganser,chen2023self,du2024electroencephalographic}-aim to alleviate the scarcity of EEG data by increasing data diversity, they often fail to reduce label ambiguity. Second, most domain adaptation frameworks rely on static or shallow distribution alignment techniques, which are insufficient to model the evolving discrepancies between marginal and conditional distributions in high-dimensional EEG feature spaces. This limitation results in poor generalization and unstable adaptation performance across heterogeneous subjects. Third, existing contrastive learning and similarity-based methods typically depend on temporally synchronized experimental protocols or pseudo-label assumptions to construct positive and negative sample pairs. These dependencies implicitly introduce supervisory signals, contradicting the assumption of fully unsupervised learning and limiting their applicability in real-world, asynchronous, or spontaneous emotional scenarios. Moreover, in the absence of reliable guidance, these methods struggle to capture fine-grained semantic boundaries in the target domain, leading to increased class ambiguity and reduced discriminability.

To overcome these limitations, we propose a novel domain adaptation framework for cross-subject EEG emotion recognition, which systematically tackles individual variability, semantic structure modeling, and distribution alignment. The main contributions of this work are as follows:

(1) Same-Subject Same-Trial Mixup (SS-Mix): Inspired by the Mixup augmentation strategy\cite{berthelot2019mixmatch}, we design an intra-subject, intra-trial sample mixing mechanism to enhance data diversity while preserving individual-specific traits. This strategy effectively mitigates the ambiguity between subject-specific features and emotion labels.

(2) Dynamic Distribution Alignment (DDA) in RKHS: We construct a unified kernel mean embedding framework that jointly aligns marginal and conditional distributions in a shared RKHS space. This objective is formulated as a multi-objective optimization problem to adaptively balance subject-level (global) and emotion-level (semantic) alignment. A dynamic confidence-based mechanism is further introduced to progressively filter high-confidence pseudo-labeled samples, improving conditional alignment reliability and transfer robustness.

(3) Dual-Domain Similarity Consistency Learning (DSCL): We propose a structure-aware constraint that enforces pairwise similarity consistency across both source and target domains. This approach enables the model to effectively capture fine-grained semantic boundaries without relying on temporal synchronization. Consequently, it enhances generalization capabilities in complex and unlabeled emotional scenarios.
Notations and descriptions used in this paper is shown in Table\ref{tab1}.
\begin{table}[h!]
	\begin{center}
		\caption{Notations and descriptions used in this paper.}\label{tab1}
		\setlength{\tabcolsep}{1mm}
		\begin{tabular}[t]{l|c}
			\hline
			Notation & Description \\
			\hline
			%S$\setminus$T & Source$\setminus$Target domain \\
			$D_s=\left\{ x_{s}^{l},y_{s}^{l} \right\} _{i=1}^{n_s}$ &Source domain\\
			$D_t=\left\{ x_{t}^{u} \right\} _{j=1}^{n_t}$&Target domain \\
			$ x_{s}^{l}/x_{t}^{u}$ & Source$/$Target domain samples\\
			$ y_{s}^{l} $ & Ground truth labels in the source domain\\
			$n_s$&  Number of source domain samples\\
			$n_t$&	Number of target domain samples\\
			$f()$& Feature extractor \\
			$\hat{y}_{s}^{l}$ & Predicted labels for source domain \\
			$\hat{y}_{t}^{u}$ & Pseudo labels for target domain\\
			$K$ & Gaussian Kernel function \\
			$RKHS$ & Reproducing Kernel Hilbert Space\\
			$SGD$  & Stochastic Gradient Descent\\
			$ReLU$ & Rectified Linear unit activation function\\
			\hline
		\end{tabular}
	\end{center}
\end{table}

%\begin{figure*}
%	\centering
%	\includegraphics[width=1\linewidth]{Figure_related.png}
%	\caption{Traditional machine learning models are trained in isolation on separate datasets, without benefiting from shared knowledge. Transfer learning improves efficiency by reusing the knowledge from a model trained on a source dataset and applying it to a different target dataset. Within transfer learning, domain adaptation specifically addresses domain shift, a scenario where the data distributions of $D_s$ and $D_t$ are misaligned. By aligning these distributional differences, domain adaptation methods enable models to perform robustly on $D_t$, even with minimal labeled data.} \label{fg:related}
%\end{figure*} 

\section{METHODOLOGY}
\begin{figure*} 
	\centering
	\includegraphics[width=1\linewidth]{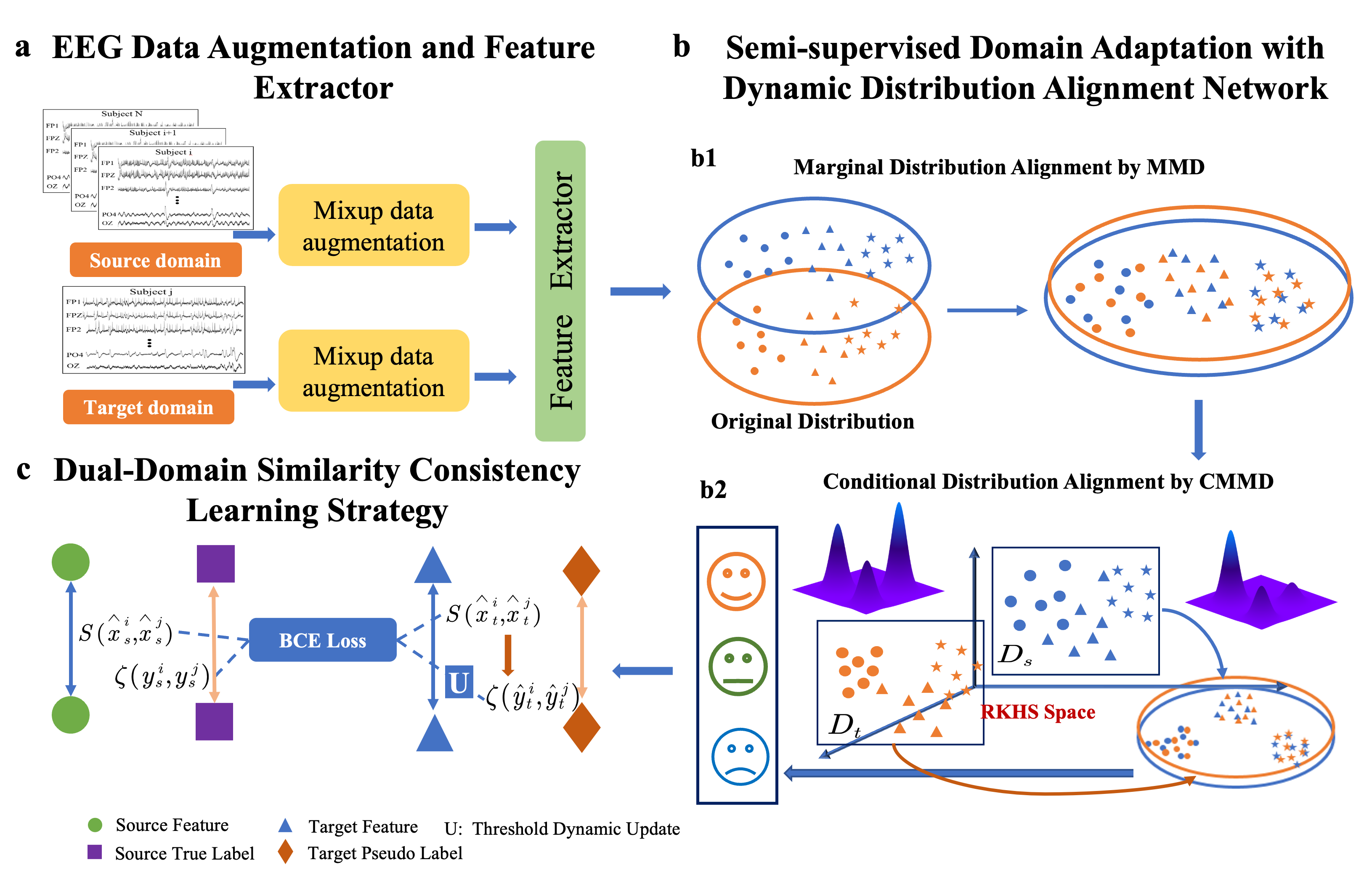}
	\caption{The flowchart of the proposed SDC-Net framework.}\label{fg:framework}
\end{figure*} 

\subsection{Data augmentation and Feature Extractor}
\subsubsection{EEG data augmentation Based SS-Mix} 
The Mixup method performs data augmentation by creating convex combinations of pairs of samples, which extends the distribution space of the training data to a certain extent and enhances the model's generalization capability. This simple yet effective method has been proven to possess unique advantages in improving model robustness and reducing overfitting.

The mathematical expression for Mixup can be represented as follows:
\begin{equation}\label{MMD1}
	\widetilde{x}=\omega x_i+\left( 1-\omega \right) x_j
\end{equation}
\begin{equation}\label{MMD1}
	\widetilde{y}=\omega y_i+\left( 1-\omega \right) y_j
\end{equation}
where $\left( x_i,y_i \right)$ and $\left( x_j,y_j \right)$ are two examples selected randomly from the training data. 

Based on the aforementioned approach, we further refined the data augmentation strategy by individually extracting and augmenting data from the same subject and the same trial, rather than augmenting all data together. This method better preserves individual differences across trials within each subject and ensures that the augmented samples maintain consistency and physiological relevance.

Specifically, we proposed a trial-wise augmentation strategy, termed SS-Mix, where each subject's dataset was segmented into individual trials, and the Mixup method was applied separately within each trial. In this setting, sample pairs were augmented as follows:
\begin{equation}\label{mixup_individual1}
	\tilde{x}_{trial}=\omega x_{trial}^{i}+\left( 1-\omega \right) x_{trial}^{j}
\end{equation}
\begin{equation}\label{mixup_individual2}
	\tilde{y}_{trial}=\omega y_{trial}^{i}+\left( 1-\omega \right) y_{trial}^{j}
\end{equation}
Here, $x_{trial}^{i}$ and $ x_{trial}^{j}$ are two randomly selected samples within the same trial, $ y_{trial}^{i}$ and $ y_{trial}^{j}$ are their corresponding labels, where $\omega$ is a random mixing coefficient drawn from a Beta distribution over the interval $(0, 1)$. Unlike global data augmentation, this method ensures the consistency of data within trials while further enhancing the preservation of individual differences during the augmentation process. This approach not only expands the distributional space of the data but also minimizes potential physiological and psychological differences between subjects that could interfere with model training. Consequently, it enhances the model's generalization capability and robustness in cross-subject emotion recognition tasks.

\subsubsection{Feature Extractor}
To extract informative representations from raw EEG signals, we first applied short-time Fourier transform (STFT) to decompose the data into five canonical frequency bands: delta ($\delta$), theta ($\theta$), alpha ($\alpha$), beta ($\beta$), and gamma ($\gamma$). For each band, we estimated the probability density function (PDF) of the amplitude distribution and computed its differential entropy (DE):
\begin{equation}\label{entropy}
	H(x) = -\int p(x)\log p(x) dx,
\end{equation}
where $p(x)$ denotes the PDF. DE provides a compact measure of the information content within each band. Concatenating the DE features across all channels and frequency bands yields a spectral–spatial feature vector of dimension $N_{cf}=c\times f$ ($c$: channels, $f=5$: frequency bands), which serves as the input to the network.

\subsection{Dynamic Distribution Alignment in RKHS}

To address the distributional discrepancy between source domain $D_s$ and target domain $D_t$ in EEG emotion recognition, we propose a unified framework named Dynamic Distribution Alignment in RKHS. This method performs joint alignment of MPD and CPD in a shared RKHS. The key idea is to unify marginal and conditional alignment into a single kernel mean embedding (KME)-based framework. The detailed derivations of Dynamic Distribution Alignment in RKHS are provided in the Supplementary Information (Section~S1).

\subsubsection{Unified Alignment Framework}
We adopt a unified kernel mean embedding (KME) framework to align both marginal and conditional distributions in the RKHS. 
Formally, the alignment loss is defined as
\begin{equation}
	\mathcal{L}_{\text{align}}
	= \mathbb{E}_{y\sim Y}\|\mu_{P(x|y)}-\mu_{Q(x|y)}\|_{\mathcal{H}}^2,
\end{equation}
where $\mu_{P(x|y)}$ and $\mu_{Q(x|y)}$ denote the conditional embeddings of the source and target domains. Detailed derivations and special cases (MMD, CMMD) are provided in the Supplementary Information (Section~S1.1).

\subsubsection{Alignment of Marginal Distributions in RKHS}
To reduce global distribution shift, we employ the Maximum Mean Discrepancy (MMD) between source and target domains:
\begin{equation}
	\mathcal{L}_{\text{MMD}} = \big\| \tfrac{1}{n}\!\sum_{i=1}^n \varPhi(x_i^s) 
	- \tfrac{1}{m}\!\sum_{j=1}^m \varPhi(x_j^t) \big\|_{H}^2,
	\label{eq:MMD}
\end{equation}
where $\varPhi(\cdot)$ denotes the feature mapping into RKHS. 
The empirical estimation and multi-kernel extension of MMD are derived in the Supplementary Information (Section~S1.2).

\subsubsection{Alignment of Conditional Probability Distributions in RKHS}
To ensure robust conditional distribution alignment, we first introduce a dynamic confidence-based selection mechanism for pseudo-labels in the target domain. Specifically, for each target sample $\mathbf{x}_i^t$, let $\hat{\mathbf{y}}_i^t$ denote its predicted label distribution. If the confidence $\max(\hat{\mathbf{y}}_i^t) \geq \tau$, the pseudo-label is accepted; otherwise, the sample is excluded from CMMD computation. Formally:
\begin{equation}
	\hat{D}_t = \left\{ \Gamma(p_i^u, \tau) \cdot \hat{y}_t^u \right\}_{i=1}^{N_u}, \quad \tau \in [0,1]
	\label{eq:pseudo-label}
\end{equation}
The threshold $\tau$ is gradually increased during training, allowing more reliable samples to be incorporated over time. Consequently, the final CMMD loss is computed only on high-confidence samples from $\hat{D}_t$.

Beyond marginal alignment, we further enforce semantic consistency by aligning class-conditional distributions across domains. The CMMD loss is defined as
\begin{equation}
	\mathcal{L}_{\text{CMMD}}=\sum_{c=1}^C \|\mu_{P(x|y=c)}-\mu_{Q(x|y=c)}\|_{\mathcal{H}}^2,
	\label{eq:CMMD}
\end{equation}
where $\mu_{P(x|y=c)}$ and $\mu_{Q(x|y=c)}$ denote the mean embeddings of source and target samples for class $c$ in RKHS. Detailed derivations of CMMD are provided in the Supplementary Information (Section~S1.3).

\begin{figure*} 
	\centering
	\includegraphics[width=0.95\linewidth]{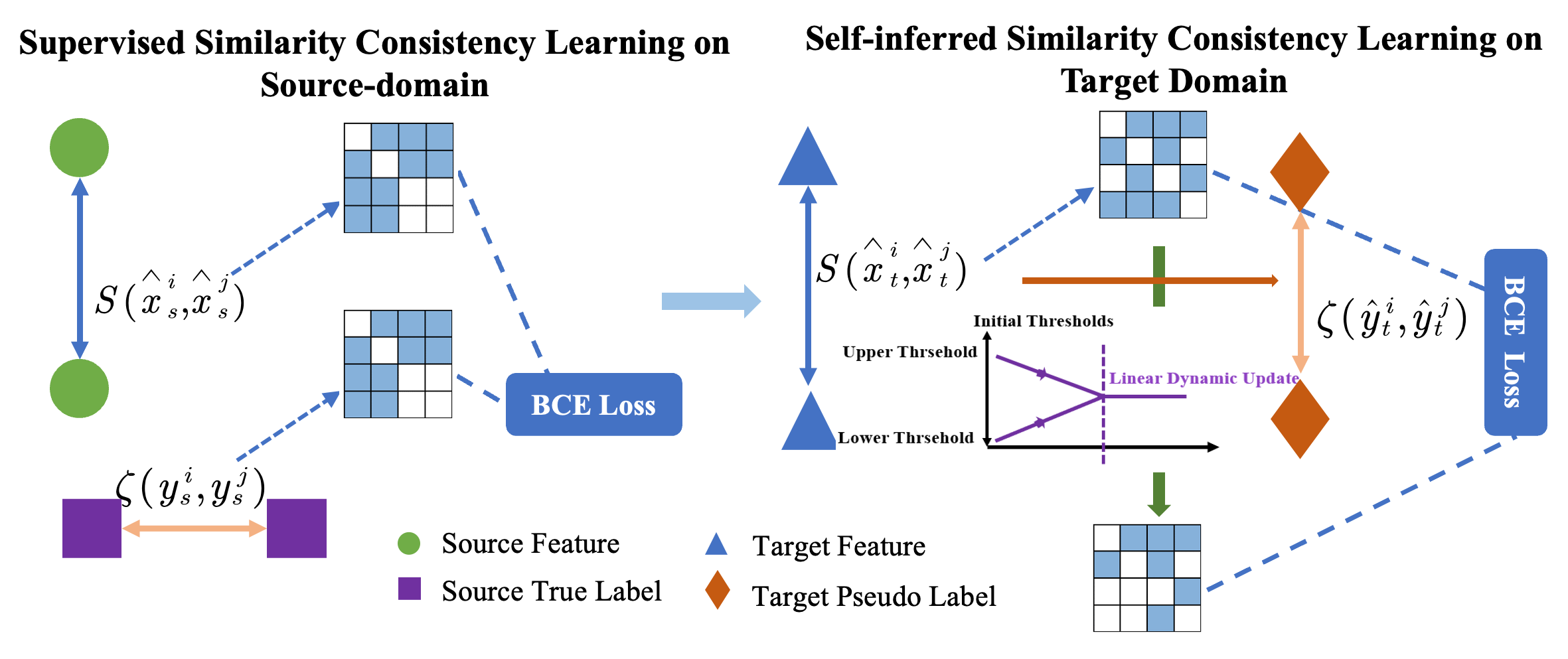}
	\caption{An illustration of the proposed DSCL strategy. The left part shows supervised similarity consistency learning on the source domain using ground-truth labels. The right part depicts self-inferred similarity consistency learning on the target domain based on pseudo-labels and dynamic similarity thresholds. Pairwise similarity is enforced through BCE loss in both domains to promote intra-class compactness and inter-class separation.}\label{fg:consistency}
\end{figure*}
\subsection{Dual-Domain Similarity Consistency Learning Strategy}

To promote discriminative and transferable feature representations for EEG signals, we propose a DSCL strategy. This module encourages the learned feature space to maintain pairwise semantic consistency—ensuring that samples from the same class are closer, and those from different classes are further apart. In order to achieve it, pairwise similarity learning is applied in both the source and target domains, as illustrated in Fig.\ref{fg:consistency}.

\subsubsection{Supervised Similarity Consistency on Source Domain}

In the source domain $D_s$, ground-truth labels are available, allowing us to supervise the model’s feature similarity predictions. Given two source samples $x_s^i$ and $x_s^j$ with labels $y_s^i$ and $y_s^j$, we define their true semantic similarity using a binary indicator:

\begin{equation}
	\zeta(y_s^i, y_s^j) = \begin{cases}
		1, & \text{if } y_s^i = y_s^j \\
		0, & \text{otherwise}
	\end{cases}
\end{equation}

Their feature similarity is computed by cosine similarity over the normalized embeddings $\widehat{x}_s^i$ and $\widehat{x}_s^j$:

\begin{equation}
	S(\widehat{x}_s^i, \widehat{x}_s^j) = \frac{\widehat{x}_s^i \cdot \widehat{x}_s^j}{\|\widehat{x}_s^i\| \cdot \|\widehat{x}_s^j\|}, \quad S' = \frac{S + 1}{2}
\end{equation}

We map $S$ to the $[0,1]$ interval as $S'$, allowing it to be compared with the binary label using binary cross-entropy (BCE) loss. The supervised similarity loss is:

\begin{equation}
	\mathcal{L}_{ps} = \frac{1}{N_s(N_s - 1)} \sum_{i \neq j} l\left(\zeta(y_s^i, y_s^j), S'(\widehat{x}_s^i, \widehat{x}_s^j)\right)
	\label{eq:pairwise_source}
\end{equation}
where $l(\cdot)$ is the BCE loss. This objective enforces semantic structure preservation within the source domain, enhancing intra-class compactness and inter-class separation.

\subsubsection{Self-Inferred Similarity Consistency on Target Domain}
In the target domain $D_t$, labels are unavailable. To still enable pairwise semantic learning, we utilize pseudo-labels $\hat{y}_t$ generated from the model itself using the confidence-based filtering strategy.

The key idea is to \textit{infer} pairwise relationships based on cosine similarity. For target feature embeddings $\widehat{x}_t^i$ and $\widehat{x}_t^j$, we define:

\begin{equation}
	\zeta(\hat{y}_t^i, \hat{y}_t^j) =
	\begin{cases}
		1, & \text{if } S(\widehat{x}_t^i, \widehat{x}_t^j) \geq \tau_{pu} \\
		0, & \text{if } S(\widehat{x}_t^i, \widehat{x}_t^j) < \tau_{pl}
	\end{cases}
\end{equation}

Here, $\tau_{pu}$ and $\tau_{pl}$ represent dynamic upper and lower thresholds used to determine confident positive or negative pairs, respectively. Feature similarities within the ambiguous region $[\tau_{pl}, \tau_{pu}]$ are excluded to avoid noisy supervision. These thresholds are linearly adjusted during training to gradually include more pairs as the model becomes more confident (visualized in Fig.\ref{fg:consistency}).

Only confident pairs $(i, j) \in \mathcal{P}$ are selected for learning, and the target-domain loss is:

\begin{equation}
	\mathcal{L}_{pt} = \frac{1}{|\mathcal{P}|} \sum_{(i,j) \in \mathcal{P}} l\left(\zeta(\hat{y}_t^i, \hat{y}_t^j), S'(\widehat{x}_t^i, \widehat{x}_t^j)\right)
	\label{eq:pairwise_target}
\end{equation}

\subsection{Loss Function and Training Procedure of SDC-Net}

To enable robust domain adaptation under fully unlabeled target conditions, the proposed SDC-Net is optimized with a unified objective comprising five loss components: the classification loss on the source domain ($\mathcal{L}_{Ds}$), marginal and conditional distribution alignment losses ($\mathcal{L}_{mmd}$ and $\mathcal{L}_{cmmd}$), and pairwise similarity losses on both the source ($\mathcal{L}_{ps}$) and target ($\mathcal{L}_{pt}$) domains. These components are integrated into the following total loss:

\begin{equation}\label{total-loss}
	\mathcal{L} = \mathcal{L}_{Ds} + \alpha \mathcal{L}_{mmd} + \beta \mathcal{L}_{cmmd} + \beta \mathcal{L}_{pt} + \lambda \mathcal{L}_{ps}
\end{equation}

The classification loss $\mathcal{L}_{Ds}$ supervises the prediction of labeled source domain samples using cross-entropy, and is defined as:

\begin{equation}\label{loss:ce}
	\mathcal{L}_{Ds} = -\frac{1}{B_L} \sum_{i=1}^{B_L} \sum_{c=1}^{C} y_s^l(i, c) \log p_s^l(i, c)
\end{equation}
where $B_L$ is the batch size, $C$ denotes the number of emotion classes, $y_s^l$ presents the one-hot ground-truth label, and $p_s^l$ means the predicted probability distribution.

To bridge the distributional gap between domains, we employ both marginal and conditional alignment losses. The marginal distribution alignment is performed via MMD, while conditional alignment is handled through CMMD, which utilizes high-confidence pseudo-labeled target samples. These two components guide the feature space toward domain invariance while preserving emotion-specific semantics.

In addition, to capture semantic structure, we impose similarity consistency constraints. The $\mathcal{L}_{ps}$ loss leverages source labels to model intra-class and inter-class relations, while $\mathcal{L}_{pt}$ promotes structural consistency on the target domain using high-confidence pseudo-labels. These losses enforce that samples from the same emotion class remain close in the latent space, even across domains.

To balance the influence of each loss component throughout training, we adopt a dynamic weighting strategy. The coefficient $\alpha$ begins with a high value to prioritize marginal alignment and is gradually reduced to emphasize semantic alignment in later stages. The coefficient $\beta$ is adjusted based on the classification loss via a step function:

\begin{equation}\label{beta}
	\beta = \varepsilon(\rho_0 - \mathcal{L}_{Ds}) + \frac{1}{2} \varepsilon(\mathcal{L}_{Ds} - \rho_0) \varepsilon(\rho_1 - \mathcal{L}_{Ds})
\end{equation}
where $\varepsilon(\cdot)$ is the Heaviside step function, and $\rho_0$, $\rho_1$ are two predefined thresholds.

To further enhance semantic modeling on the target domain, the weight $\lambda$ for the unsupervised pairwise loss $\mathcal{L}_{pt}$ increases linearly with training epochs:

\begin{equation}\label{lambda}
	\lambda = \frac{2e}{\text{epochs}}
\end{equation}

This progressive adjustment ensures that SDC-Net shifts from global alignment to finer semantic refinement as training evolves, ultimately improving generalization performance in fully unsupervised cross-subject EEG emotion recognition tasks. The overall learning process is provided in Algorithm S1 of the Supplementary Information.

\section{Experiments}\label{sec:experimentresults}
\subsection{Emotion datasets}
To demonstrate the effectiveness of SDC-Net for cross-subject EEG-based emotion recognition, we conduct experiments on three public benchmark datasets: SEED~\cite{zheng2015investigating}, SEED-IV~\cite{8283814}, and FACED~\cite{chen2023large}. SEED and SEED-IV contain EEG signals from 15 subjects recorded with a 62-channel NeuroScan system, induced by film clips designed to evoke three (positive, neutral, negative) and four (happiness, sadness, neutral, fear) emotions, respectively. FACED involves 123 participants with 32-channel recordings and supports both nine-class discrete emotion classification and a binary positive/negative task. 

\subsection{Experiment Setting and  Implementation Details}
We evaluated SDC-Net on the SEED and SEED-IV datasets using two widely adopted cross-validation protocols: 
(1) \textbf{Cross-subject single-session leave-one-subject-out}, where one subject was used as the target domain $D_t$ and the others as the source domain $D_s$; 
(2) \textbf{Cross-subject cross-session leave-one-subject-out}, where one subject’s entire session was treated as $D_t$ and the remaining sessions as $D_s$. 
These protocols provide a rigorous assessment of generalization across both subjects and sessions. Further procedural details are provided in the Supplementary Information (Section S2.1).

The feature extractor consists of two fully connected layers (310 $\to$ 64 $\to$ 64) with ReLU activations and dropout (0.25). 
Training was performed for 200 epochs with SGD (momentum = 0.9), batch size = 32, and initial learning rates in $\{0.001,0.01\}$. 
Additional hyperparameter specifications (e.g., weight decay, random seed, threshold schedules, and dynamic coefficients) are reported in the Supplementary Information (Section S2.2-S2.4).
\section{Result}\label{sec:discussion}
\subsection{Experimental Results}
\subsubsection{Cross-subject single-session leave-one-out-subject-out cross-validation Results}
In Tables\ref{tab2} and \ref{tab3}, we conducted a comprehensive evaluation of various representations on the SEED and SEED-IV datasets using the leave-one-subject-out cross-validation method with a cross-subject single-session protocol. Our method demonstrated significant performance advantages on the SEED dataset, achieving an accuracy of 91.85\%$\pm$ 05.98\%. Similarly, on the SEED-IV dataset, our method exhibited competitive accuracy of 74.88\%$\pm$10.47\%. All results are reported as mean$\pm$ standard deviation over test subjects. These results strongly validate the substantial performance improvements achieved by our method on both datasets, particularly the remarkable accuracy of 91.85\%$\pm$ 05.98\% on the SEED dataset, surpassing the industry average and demonstrating notable potential in the field of emotion recognition tasks. These findings provide compelling evidence supporting the effective application of our method in real-world scenarios. As shown in Table S1 (Supplementary Information (Section S3.1)), the SDC-Net model was evaluated on the FACED dataset for both binary classification (FACED-2) of positive and negative emotions and nine-class emotion classification (FACED-9), achieving accuracies of 75.2\%$\pm$8.46\% and 42.4\%$\pm$6.55\%, respectively. Compared to the baseline model DE$\pm$SVM, it achieved improvements for the binary and nine-class tasks, respectively, highlighting the superior performance of our model, particularly in handling more fine-grained emotion classification tasks.

\subsubsection{Cross-subject cross-session leave-one-out-subject-out cross-validation results}
Another crucial consideration for emotion brain-computer interfaces is the substantial variability observed among different subjects across various sessions. The evaluation approach of cross-subject and cross-session represents a significant challenge for the effectiveness of models in EEG-based emotion recognition tasks. To further validate this detection approach, which aligns more closely with real-world application scenarios, we conducted experiments and obtained outstanding three-class classification performance on the SEED dataset, achieving an accuracy of 82.22\%$\pm$04.68\% (see Table\ref{tab_combined}). Additionally, on the SEED-IV dataset, our model achieved a four-class accuracy of 68.84\%$\pm$08.05\% (see Table\ref{tab_combined}). Compared to existing research, the proposed SDC-Net method demonstrated industry-leading performance with a smaller standard deviation. These results indicate that the proposed SDC-Net method exhibits excellent stability and generalization capabilities in handling subject and session differences.

\begin{table}
	\centering
	\caption{The performance of representation methods on SEED datasets using Cross-Subject Single-Session Leave-One-Subject-Out Cross-Validation}
	\setlength{\tabcolsep}{1.8mm}
	\renewcommand{\arraystretch}{1.3}
	\label{tab2}
	\begin{tabular}{c|c|c|c}
		\hline
		Method & Acc(\%) & Method & Acc(\%) \\
		\hline
		RGNN\cite{zhong2020eeg} & 85.30$\pm$06.72 & BiHDM\cite{li2020novel} & 85.40$\pm$07.53 \\\hline
		JDA-Net\cite{li2019domain} & 88.28$\pm$11.44 & DA-CapsNet\cite{liu2024capsnet} &84.63$\pm$09.09 \\\hline
		MS-MDA\cite{chen2021ms} & 89.63$\pm$06.97 & WGAN-GP\cite{li2019regional} & 87.10$\pm$07.10\\\hline
		%MSFR-GCN\cite{pan2023msfr} & 86.78$\pm$05.40 & \textbf{SDC-Net}  & 91.85$\pm$05.98                                \\\hline
		DGGN\cite{gu2023domain} & 83.84$\pm$10.26 & EPNNE\cite{zhang2024evolutionary} & 89.10$\pm$03.60\\\hline
		DC-ASTGCN\cite{yang2024dc} & 80.65$\pm$08.46 & MS-FRAN\cite{li2023ms} & 85.61$\pm$06.55\\\hline
		SDDA\cite{li2022dynamic} & 91.08$\pm$07.70 & CU-GCN\cite{gao2024graph} & 87.10$\pm$05.44\\\hline
		DAPLP \cite{zhong2025unsupervised} & 89.44$\pm$04.89 & DS-AGC \cite{ye2024semi} & 86.38$\pm$07.25\\\hline
		PR-PL\cite{zhou2023pr} & 85.88$\pm$09.36 & PLMSDANet\cite{ren2024semi} & 84.21$\pm$12.34 \\\hline
		CLISA\cite{shen2022contrastive} & 86.40$\pm$06.40 & CL-CS\cite{hu2025cross} & 88.30$\pm$08.90 \\\hline
		ST-SCGNN\cite{pan2023st} & 85.90$\pm$04.90 & \textbf{SDC-Net}  & \textbf{91.85$\pm$05.98}\\\hline
	\end{tabular} 
\end{table}
\begin{table}
	\centering
	\caption{The performance of representation methods on SEED-IV datasets using Cross-Subject Single-Session Leave-One-Subject-Out Cross-Validation}
	\setlength{\tabcolsep}{1.8mm}
	\renewcommand{\arraystretch}{1.3}
	\label{tab3}
	\begin{tabular}{c|c|c|c}
		\hline
		Method & Acc(\%) & Method & Acc(\%) \\
		\hline
		DGCNN\cite{song2018eeg} & 68.73$\pm$08.34 & MS-STM\cite{li2019multisource} & 61.41$\pm$09.72 \\\hline
		MS-ADRT\cite{jiang2023generalization} & 68.98$\pm$06.80  & MS-MDA\cite{chen2021ms} & 59.34$\pm$05.48 \\\hline
		WGAN-GP\cite{li2019regional} & 60.60$\pm$15.76 & JDA-Net\cite{li2019domain}  & 70.83$\pm$10.25\\\hline
		ST-SCGCN\cite{pan2023st}  & 76.37$\pm$05.77 &CU-GCN\cite{gao2024graph} & 74.50$\pm$07.88 \\\hline
		DAPLP \cite{zhong2025unsupervised} & 74.57$\pm$06.18 & DS-AGC \cite{ye2024semi} & 66.00$\pm$07.93\\\hline
		MSFR-GCN\cite{pan2023msfr} & 73.43$\pm$07.32 & \textbf{SDC-Net}  & \textbf{74.88$\pm$10.47}\\\hline
		
	\end{tabular}
\end{table}
\begin{table}[t]
	\centering
	\caption{The performance of shared methods on SEED and SEED-IV datasets using Cross-Subject Cross-Session Leave-One-Subject-Out Cross-Validation}
	\setlength{\tabcolsep}{2mm}
	\renewcommand{\arraystretch}{1.3}
	\label{tab_combined}
	\begin{tabular}{c|c|c}
		\hline
		Method & SEED (Acc\%)  & SEED-IV (Acc\%) \\
		\hline
		RF\cite{breiman2001random} & 69.60$\pm$07.64 & 50.98$\pm$09.20 \\\hline
		KNN\cite{coomans1982alternative} & 60.66$\pm$07.93 & 40.83$\pm$07.28 \\\hline
		SVM\cite{suykens1999least} & 68.15$\pm$07.38 & 51.78$\pm$12.85 \\\hline
		TCA\cite{pan2010domain} & 64.02$\pm$07.96 & 56.56$\pm$13.77 \\\hline
		CORAL\cite{sun2016return} & 68.15$\pm$07.83 & 49.44$\pm$09.09 \\\hline
		SA\cite{fernando2013unsupervised} & 61.41$\pm$09.75 & 64.44$\pm$09.46 \\\hline
		GFK\cite{li2018cross} & 66.02$\pm$07.59 & 45.89$\pm$08.27 \\\hline
		DANN\cite{JMLR:v17:15-239} & 81.08$\pm$05.88 & 54.63$\pm$08.03 \\\hline
		\textbf{SDC-Net} & \textbf{82.22$\pm$04.68} & \textbf{68.84$\pm$08.05} \\\hline
	\end{tabular}
\end{table}
\subsection{Confusion matrices}
To further analyze classification behavior, we compare the confusion matrices of four representative models: DA-CapsNet, PLMSDANet, LGDAAN-Net, and our proposed SDC-Net. The detailed results are provided in Figure~S1 of the Supplementary Information (Section S3.2). The confusion matrices provide insight into how well each model classifies inputs into three categories: Negative, Neutral, and Positive. The diagonal elements represent correct predictions, while off-diagonal elements correspond to misclassifications. Our proposed SDC-Net model outperforms the other models, particularly in classifying the Neutral class with an accuracy of 92.45\%, which is the highest among all models. 

Additionally, SDC-Net achieves 95.7\% accuracy in the Positive class and 87.23\% in the Negative class. The confusion between classes is minimal, with only 5.25\% of neutral instances being misclassified as negative and 2.53\% of positive instances being misclassified as negative. The low misclassification rates in SDC-Net suggest that our model effectively captures subtle differences between sentiment classes, particularly between negative and neutral sentiments, where other models faltered. It achieves the best accuracy in both the Neutral and Positive classes and exhibits significantly lower misclassification rates compared to the other models. This indicates that SDC-Net provides a more nuanced understanding of sentiment, making it particularly effective for tasks that require fine-grained sentiment analysis. In contrast, models such as LGDAAN-Net, while demonstrating strong performance in classifying positive samples, exhibit notable deficiencies in differentiating between negative and neutral sentiments. This limitation significantly undermines their effectiveness in comprehensive sentiment classification tasks.
\subsection{Ablation Study}
To evaluate the contribution of each module in the proposed SDC-Net, we performed ablation experiments by systematically removing six key components. Table~\ref{tab6} summarizes the performance under each setting in terms of average classification accuracy. First, domain alignment mechanisms were found to be essential. Removing of the MMD component resulted in a significant drop in performance (86.82\%), indicating its crucial role in reducing marginal distribution discrepancy between source and target domains. Similarly, excluding CMMD led to a moderate performance decline (89.91\%), showing that conditional alignment further refines domain adaptation, though its impact is secondary to MMD. Second, the effect of DSCL was also prominent. Supervised similarity consistency on source domain contributes to extracting discriminative features from labeled source data, and its removal decreased the accuracy to 88.99\%. More critically, excluding self-inferred similarity consistency leanring on target domain reduced performance to 86.99\%, emphasizing its importance in modeling intra-class similarity in the unlabeled target domain, thus supporting better generalization. Third, SS-Mix based data augmentation improved generalization by synthesizing diverse EEG trials. Its removal caused a noticeable decrease in Acc (87.99\%), suggesting its role in mitigating overfitting and increasing sample diversity. Finally, the pseudo-confidence mechanism, which filters unreliable pseudo-pairs, slightly improved the mean accuracy and significantly reduced variance. Without this mechanism, the model still achieved 89.66\%, but with increased performance fluctuation (standard deviation 7.18), indicating its stabilizing effect on target domain predictions.

%In summary, MMD and target similarity consistency learning are the most influential modules in our framework. The remaining components also enhance the model’s robustness and consistency, contributing to the superior performance of SDC-Net across cross-subject EEG emotion recognition tasks.
\begin{table}
	\centering
	\caption{Performance of the SDC-Net model in the ablation study}\label{tab6}
	\setlength{\tabcolsep}{2mm}
	\renewcommand{\arraystretch}{1.3}
	\begin{tabular}{c|c}
		\hline
		Ablation Experiment Strategy & Acc (\%) \\
		\hline
		without-SS-Mix            & 87.99 $\pm$ 5.77 \\
		without-MMD                & 86.82 $\pm$ 5.86 \\
		without-CMMD               & 89.91 $\pm$ 5.03 \\
		without-similarity consistency on $D_s$   & 88.99 $\pm$ 6.06 \\
		without-similarity consistency on $D_t$  & 86.99 $\pm$ 5.65 \\
		without-pseudo-confidence  & 89.66 $\pm$ 7.18 \\
		\hline
		\textbf{SDC-Net}           & \textbf{91.85 $\pm$ 5.98} \\
		\hline
	\end{tabular}
\end{table}
\begin{figure}[t]
	\centering
	\includegraphics[width=0.9\linewidth]{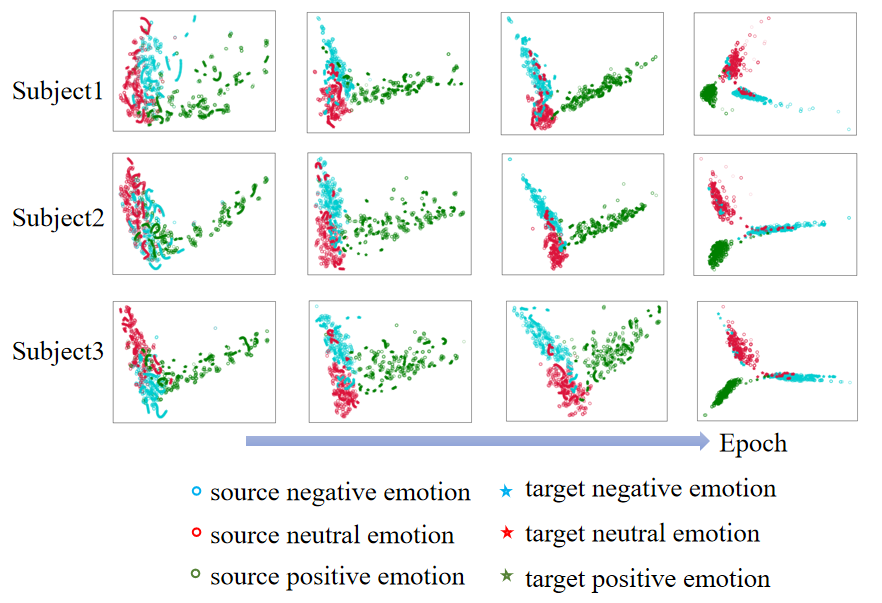}
	\caption{t-SNE visualization of source and target domain emotion representations for three subjects across training epochs.}
	\label{fig:tsne}
\end{figure}
\begin{figure*} 
	\centering
	\includegraphics[width=.9\linewidth]{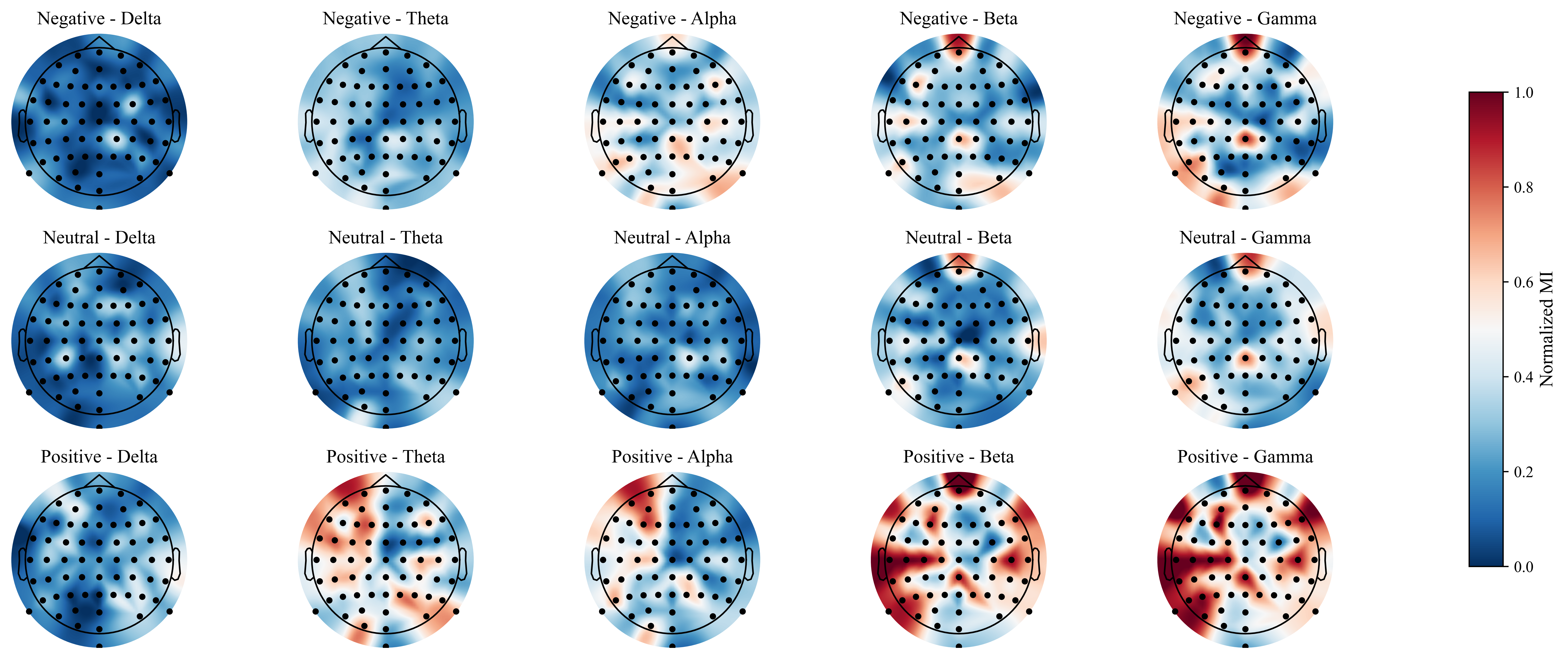}
	\caption{Topographic analysis of the mutual information between EEG frequency-band features and model predictions across emotional states.}\label{fg:topographic}
\end{figure*} 
\subsection{Visualization of Domain Alignment via t-SNE}
To qualitatively assess the effectiveness of domain alignment, we visualized the feature distributions of source and target domains using t-SNE at different training stages (Fig.\ref{fig:tsne}). In these plots, the source domain ($D_s$) is represented by circles and the target domain ($D_t$) by stars, with emotion categories indicated by color (blue: negative, red: neutral, green: positive). At the early training stage, features from $D_s$ and $D_t$ are poorly aligned. The target samples, especially for Subject 1 and Subject 2, are scattered and show significant overlap across emotion classes, indicating large domain discrepancy. Positive target samples (green stars) frequently overlap with other categories, suggesting the initial failure in feature alignment. As training progresses, domain alignment improves notably. For Subject 2 and 3, the target domain emotion clusters begin to align with those of the source domain. Notably, negative and positive emotions become more separable, suggesting that MMD and CMMD modules effectively minimize marginal and conditional distribution gaps. By the final training stage, most emotion clusters exhibit clear separation in both domains. Subject 3 shows the most consistent alignment across categories, while Subject 1 retains partial overlap in neutral emotions. This aligns with known challenges in classifying neutral emotional states, which often lie closer to the decision boundaries due to their ambiguous EEG patterns.

%Overall, the t-SNE visualizations provide intuitive support for the proposed framework’s ability to learn emotion-discriminative and domain-invariant features. Although some overlap remains—particularly for neutral states, the progressive clustering over epochs confirms the effectiveness of the domain adaptation strategy.

\subsection{Negative Transfer Results}
This study evaluates the effectiveness of the proposed SDC-Net framework in alleviating negative transfer in fully unsupervised cross-subject EEG emotion recognition across 45 experimental tasks. Negative transfer is defined as a classification accuracy lower than 33.3\% on the SEED dataset or 25\% on the SEED-IV dataset. As shown in Table\ref{tab7}, SDC-Net achieved zero instances of negative transfer, demonstrating a clear advantage over conventional transfer learning approaches.

The robustness of SDC-Net in avoiding negative transfer can be attributed to three key components:(1)SS-Mix, which augments intra-subject data while preserving individual-specific characteristics, reducing ambiguity between subject identity and emotion labels;(2) Dynamic Distribution Alignment in RKHS, which jointly aligns marginal and class-conditional distributions using a unified kernel mean embedding framework, while adaptively filtering high-confidence pseudo-labels to ensure alignment quality;(3) DSCL, which imposes structural constraints on pairwise similarities across domains, promoting semantic consistency without relying on temporal alignment. To sum up, these innovations enable SDC-Net to dynamically adjust to distribution shifts, suppress noisy pseudo-labels, and maintain semantic integrity across subjects, thereby ensuring stable and reliable transfer performance in complex, label-free EEG emotion recognition scenarios.
\begin{table}[ht]
	\centering
	\caption{Number of Subjects Exhibiting Negative Transfer on SEED and SEED-IV Datasets}\label{tab7}
	\setlength{\tabcolsep}{5mm}
	\begin{tabular}{c|c|c}
		\hline
		Method & SEED  & SEED-IV \\
		\hline
		PR-PL\cite{zhou2023pr} & 1 & 3 \\\hline
		MS-MDA\cite{chen2021ms} & 3 & 8 \\\hline
		PLMSDANet\cite{ren2024semi} & 0 & 1 \\\hline
		\textbf{SDC-Net} & \textbf{0} & \textbf{0} \\\hline
	\end{tabular}
\end{table}

\subsection{Topographic Analysis of Important EEG Patterns}
This study utilizes an EEG topographic mapping approach grounded in Mutual Information (MI) to evaluate the contributions of different brain regions and frequency bands to emotion classification. The preprocessed EEG feature matrix $\mathrm{\mathbf{X}} \in \mathbb{R}^{N \times 5 \times 2}$, incorporating signals from 62 electrode channels across five frequency bands, is paired with the classification probability matrix $\mathrm{\mathbf{Y}} \in \mathbb{R}^{N \times 3}$, corresponding to three emotional states: negative, neutral, and positive. By computing the MI between each EEG feature and the predicted probabilities of the emotion categories, an initial MI $MI\in R^{3\times 5\times 2}$ tensor is derived, capturing the nonlinear statistical dependencies. This tensor is then normalized via min-max scaling to the [0, 1] interval to enable fair comparison across dimensions. The normalized MI scores are then structured into a three-dimensional tensor (emotion × frequency band × channel) and visualized topographically using the standard 10–20 system for electrode positioning. A 3×5 grid layout illustrates MI distributions across the Delta (1–4 Hz), Theta (4–8 Hz), Alpha (8–13 Hz), Beta (13–30 Hz), and Gamma (30–50 Hz) bands for each emotional category. Color gradients denote the strength of feature importance, with electrode positions highlighted as black dots. This methodology effectively integrates information-theoretic metrics with neuroimaging visualization, offering intuitive insights into the spatial and spectral dynamics of emotion processing.

The results depicted in Fig.\ref{fg:topographic} indicate that the most informative EEG patterns for emotion recognition are predominantly concentrated in the beta and gamma frequency bands within the frontal and temporal regions. These findings are consistent with prior research, reinforcing the role of high-frequency oscillations in affective processing\cite{song2018eeg,harmon2010role,zhou2024eegmatch}.

\subsection{Discussion}
Addressing significant inter-subject variability in EEG data remains a central challenge in developing robust and generalizable aBCI systems. Beyond accurate classification, the ultimate goal of aBCIs is to achieve stable, interpretable, and scalable emotion measurement across users and scenarios. To this end, we propose SDC-Net, designed to improve not only recognition performance but also the consistency and robustness of emotion measurement. We conducted comprehensive evaluations on three public EEG emotion datasets-SEED, SEED-IV, and FACED-demonstrating that SDC-Net significantly outperforms state-of-the-art methods in cross-subject emotion recognition tasks, validating its effectiveness and broad applicability in affective computing and intelligent instrumentation.

One of the key innovations in SDC-Net is the SS-Mix module, which generates augmented samples through intra-trial interpolation. Compared with GAN-based data augmentation methods (e.g., GANSER\cite{zhang2022ganser} and SA-cWGAN\cite{chen2023self}), which may introduce low-quality or identity-inconsistent samples, SS-Mix enhances the modeling of intra-subject emotional variability while significantly reducing semantic noise. From a measurement perspective, this approach can be regarded as a strategy for increasing reliable sample density without introducing cross-subject artifacts, which is especially important in single-trial settings or when data is limited. To address the limitations of prior methods that rely on static distribution alignment and fixed pseudo-labels-such as BiHDM\cite{li2020novel}, MS-MDA\cite{chen2021ms}, MS-FRAN\cite{li2023ms}, DA-CapsNet\cite{liu2024capsnet}, SDDA\cite{li2022dynamic}, PR-PL\cite{zhou2023pr}, and DAPLP\cite{zhong2025unsupervised}-we propose DDA strategy in RKHS, which provides finer adaptation to inter-subject and inter-class distributional shifts. This strategy also incorporates a confidence-based pseudo-label filtering mechanism that dynamically selects target samples with well-defined semantic structures, enabling progressive alignment from easy to hard examples. Functionally, this design can be viewed as a robust calibration mechanism when labels are noisy or incomplete, significantly improving training stability and generalization. The proposed DSCL module further addresses the limitations of existing contrastive learning methods (e.g., CLISA\cite{shen2022contrastive}, CL-CS\cite{hu2025cross}) that rely on temporal synchronization and are difficult to generalize to cross-scenario measurement. DSCL infers latent structural similarity between domains and imposes corresponding consistency constraints, enabling semantic boundary learning without time alignment or pseudo-label supervision. Experimental results suggest DSCL functions as latent structural regularization, guiding the model toward learning compact and well-separated topological structures.

Ablation studies indicate that removing either the structural consistency module or the dynamic reweighting mechanism results in a 4\%–6\% drop in performance, highlighting their critical roles. Sensitivity analysis further shows that SDC-Net maintains high robustness under different kernel numbers and similarity thresholds, with performance fluctuation consistently below 2\% (detailed results are provided in Supplementary information (Section S4)). The robustness of SDC-Net is also validated through visualization and negative transfer testing. As shown in Fig.~\ref{fig:tsne}, emotional categories form well-defined and separable clusters across source and target domains. Additionally, SDC-Net demonstrates strong resistance to negative transfer, ensuring reliable emotion measurement in cross-subject settings.

\section{Conclusion}\label{sec:conclusion}
This paper proposes SDC-Net, a novel domain adaptation framework designed to address the challenge of individual variability in aBCIs for cross-subject EEG emotion recognition. The framework integrates three key innovations: (1) the Same-Subject Same-Trial Mixup data augmentation strategy; (2) dynamic distribution alignment in the RKHS; and (3) dual-domain similarity consistency learning strategy. Collectively, these components significantly enhance the capability of generalization and robustness for emotion recognition models across different subjects. Extensive experimental results on the SEED, SEED-IV, and FACED datasets demonstrate the superiority of SDC-Net over existing methods in both cross-subject and cross-session scenarios, achieving significant improvements in emotion recognition performance. 

Future work may focus on two promising directions. First, enhancing pseudo-label quality by balancing label confidence and quantity remains crucial. Adaptive confidence thresholds or robust filtering strategies could help mitigate the trade-off between discarding noisy labels and preserving class diversity \cite{zhang2021flexmatch, wang2022freematch}. Second, deploying the SDC-Net framework in real-time aBCI systems represents a crucial step toward dynamic, user-adaptive emotion measurement. Integration with online EEG acquisition platforms could enable continuous learning from streaming signals, facilitating personalized and context-aware affective computing in real-world scenarios, such as clinical monitoring or closed-loop neurofeedback instrumentation.
\balance
\bibliographystyle{IEEEtran}
\bibliography{manuscript}

\end{document}